\documentstyle[preprint,aps,eqsecnum]{revtex}
\begin{document}

\preprint{submitted to PRL}
\title
{Logarithmic Temperature Dependence of Conductivity at Half Filled Landau
 Level}

\author{D.V. Khveshchenko}
\address
{NORDITA, Blegdamsvej, 17, Copenhagen DK-2100, Denmark}

\maketitle

\begin{abstract}
\noindent
We study temperature dependence of diagonal conductivity at half
filled Landau level by means of the theory of composite fermions in the 
weakly disordered regime $(k_{F}l>>1)$. 
At low temperatures we find the leading $\log T$ correction resulting 
from interference between impurity scattering and gauge interactions
of composite fermions. 
The prefactor appears to be strongly enhanced as compared to the
standard Altshuler-Aronov term  in agreement with recent experimental
observations.

\end{abstract}
\pagebreak

Since 1990 a number of strong experimental evidencies 
of an existence of compressible metal-like 
states of two-dimensional (2D) electrons in a strong magnetic field at
even denominator fractions  was obtained \cite{exp1}.
Remarkably, the results of these experiments 
can be interpreted in terms of a simple semiclassical picture of 
spinless fermionic excitations forming a Fermi surface at $k_{F}=(4\pi n_{e})^{1/2}$ and experiencing only the difference magnetic field $\Delta B = B-4\pi qn_{e}$ in the vicinity of  $\nu={1\over 2q}$. 
The appropriate theoretical framework implied by this picture was elaborated
by Halperin, Lee, and Read \cite{HLR} on the basis of the idea of composite fermions (CF) viewed as $2q$ gauge flux quanta attached to spin-polarized electrons \cite{J}.

The mean field theory developed in \cite{HLR} allows a qualitatively successful understanding of the observations
made in \cite{exp1} in terms of a nearly Fermi liquid behavior of 
CF. However more recent experimental studies revealed features
which can not be readily explained by the picture of weakly interacting 
fermions.  

These puzzling data include a divergency of a CF effective mass extracted from measurements of magnetoresistivity at incompressible fractions $\nu={N\over {2N\pm 1}}$
converging towards $\nu=1/2$ and unusual $\sim 1/(\Delta B)^{4}$ scaling
of the corresponding Dingle plot at low temperatures \cite{exp2}.
Moreover, temperature dependence of the resistivity at $\nu=1/2$ and $3/2$ exhibits
the leading $\log T$ behavior with unusually large coefficient \cite{log}.

These striking features are not unexpectable on general grounds because 
in contrast to the case of weakly interacting fermion gas the formation
of Fermi surface at even denominator fractions results from minimization
of interaction rather than kinetic energy. Therefore one could well expect 
this system to provide a genuine 
example of 2D non-Fermi liquid (NFL).

On theoretical side, the NFL behavior stems from long-ranged retarded
gauge interactions between CF in presence of disorder \cite{HLR}. Both gauge interactions and impurity scattering of CF are more
singular than their counterparts in the original electron representation. 
So far all theoretical studies were concentrated on either effects of dynamic gauge interactions in a pure system or on a noninteracting problem of
impurity scattering of CF
which translates into a problem of static random magnetic field (RMF).
These problems arising in other contexts as well are certainly of a great interest but in a real system provided by $GaAs/Al_{x}Ga_{1-x}As$
heterojunction the number of strong Coulomb impurities (ionized donors)
is essentially equal to the number of electrons. Therefore impurities have to be
considered as an inherent element of the system and one should expect a variety
of interference effects which were extensively studied in the case of zero (or weak) magnetic field over a decade ago \cite{AA,LR}.

In the present letter we estimate the low temperature conductivity 
$\sigma_{xx}(T)$ at exactly
half filled Landau level proceeding along the lines of the previous analysis
at $B=0$ \cite{AA}.  The calculable quantity is a conductivity of CF 
which is related to the physical conductivity   as 
$\sigma^{CF}=\sigma^{2}_{xx}+\sigma^{2}_{xy}/\sigma_{xx}\approx 
{({e^2\over 2h})^2 \sigma^{-1}_{xx}}$.

In the Coulomb gauge ${div}{\vec A}=0$
the gauge interaction between CF at $\omega\tau_{tr} <<1$ and $ql<<1$
is described by the inverse of the 
matrix \cite{HLR}:
\begin{equation}
D^{-1}_{\mu\nu}(\omega, q)=
\pmatrix
{
{m\over 2\pi}{Dq^2\over {Dq^2 - i\omega}} &
-i{q\over 4\pi}
\cr
i{q\over 4\pi} & 
-i\gamma_{q}\omega + \chi_{q}q^2 \cr}
\end{equation}
where $D={1\over 2}v^{2}_{F}\tau_{tr}$ is a diffusion coefficient, $\chi_q ={1\over 12\pi m}+{1\over (4\pi)^2}V_q$ is an effective diamagnetic 
susceptibility given in terms of
 the pairwise electron potential $V_q$,
$\gamma_q ={n_e l\over k_F}$ is proportional to the CF mean free path (MFP) $l=v_F
\tau_{tr}$,
 and $m$ stands for the CF effective mass.

Depending on the structure of a concrete device
electron Coulomb interaction potential $V_q ={2\pi e^2\over q}$
can be screened by image charges induced in a ground plate.
Therefore we shall consider both cases of Coulomb and short-range
$(V_q \approx V_0 ={2\pi e^2\over \kappa}$ where $\kappa$ stands for
a screening constant) potentials.  

Below we shall see that $m$ (as well as details of the interaction potential $V_q$) drop out of final expressions
for the conductivity. In fact, our calculations do not require a detailed knowledge of the frequency dependence of CF Green function in the metallic
regime but only its momentum dependence. 

Each charged impurity placed on distance $d_s$ away
from 2D electron gas creates a scalar potential with 2D Fourier transform given by $A^{(0)}_{0}(q)={2\pi e^2\over q}e^{-qd_s}$. Due to 2D screening by gapless CF this potential gets renormalized
and also acquires a vector component corresponding to the gauge flux
located at impurity position. In the random phase approximation
the renormalized potential has a form \cite{HLR}:
\begin{equation}
A_{\mu}(q)=2\pi e^{-qd_s} 
\pmatrix
{{1\over m} \cr
{2i\over q} \cr}
\end{equation}
After averaging over positions of impurities with concentration $n_{i}$ the vector disorder appears to be 
equivalent to RMF correlated as $<B_{q}B_{-q}>=(4\pi)^2 {n_{i}}e^{-2qd_{s}}$. In what follows we put
impurity consentration equal to electron density $n_{e}$.

The RMF problem received a lot of attention
over last few years and is now believed to have localization properties
described by the unitary
random scattering ensemble \cite{AMW}. Although a total scattering rate
$1/\tau =m\int{d\theta\over 2\pi}({v_F\over q})^2 <B_{q}B_{-q}>$ where 
$q=2k_{F}\sin\theta/2$ governing a non-gauge invariant
single particle Green function 
$G_{\pm}(\epsilon, \vec p)=(\epsilon - \xi(\vec p)-\Sigma(\epsilon)
\pm{i/2\tau})^{-1}$ appears to be divergent, it is a finite momentum relaxation rate
$1/\tau_{tr}=m\int{d\theta\over 2\pi}(1-\cos\theta)
({v_F\over q})^2<B_{q}B_{-q}>=
{4\pi n_{i}\over {m}}(k_{F}d_{s})^{-1}$ which enters all physical quantities.
Therefore one can safely operate with $1/\tau$ as if it were finite \cite{AMW}.
The CF
self-energy $\Sigma(\epsilon)$ due to gauge fluctuations
depends mostly on energy $\epsilon$ and not on momentum \cite{HLR}.
It guarantees that drastic effects of gauge fluctuations on the CF Green function
resulting in a divergency of CF effective mass \cite{HLR}
do not affect our results. To put it another way, the following calculations
do no rely on validity of quasiparticle picture (although it can be justified in the $T=0$ Coulomb case with no impurities where $\Sigma^{'}(\epsilon)\sim \epsilon\log\epsilon >>\Sigma^{''}(\epsilon)\sim \epsilon$ \cite{H})
and they can be also performed by means of a more general approach of
quantum kinetic equation \cite{QBE}
generalized onto the case with impurities \cite{DVK}.

The RMF transport time $\tau_{tr}$ determines a classical value of CF
conductivity $\sigma^{CF}_{0}={e^2\over 2h}(k_{F}d_{s})>>{e^2\over h}$ found in \cite{HLR}. Note that the scalar component of (2) gives a
contribution to $\sigma^{CF}_{0}$ which is $(k_{F}d_{s})^2$ times smaller.
 
The corresponding MFP of CF $l\sim d_{s}$
is small compared with its value at zero magnetic field $l_{0}\sim 
k^{-1}_{F}(k_{F}d_{s})^3 \cite{HLR}$.
Nevertheless a metallicity parameter $k_{F}l$ can be still appreciably large
(typically, $k_{F}d_{s}\sim 15$ \cite{HLR}) which allows one to estimate corrections to $\sigma^{CF}_{0}$ by means
of a standard perturbative calculation of a linear current response \cite{AA}.

It was pointed out in \cite{HLR,KZ} that weak-localization corrections 
$\delta\sigma^{CF}_{wl}\sim {1\over \epsilon_{F}\tau_{tr}}\log T\tau_{tr}$ are negligible
because as a consequence of broken time-reversal symmetry there is no pole in the particle-particle (Cooperon) channel of the two-particle Green function. However the pole in the particle-hole channel
(diffusion) 
\begin{equation}
\Gamma(\epsilon, \omega, q)={1\over m\tau^2}
{1\over {Dq^2 -i\omega}}
\end{equation}
 is still present at
 $\omega\tau_{tr}<<1$, $ql<<1$, and
$\epsilon (\epsilon +\omega)<0$ 
since it is due to particle number conservation only \cite{AMW,F}.

A complete treatment of the CF problem requires a simultaneous account of singular gauge interactions $D_{\mu\nu}$ and impurity ladders responsible for 
the diffusion pole $\Gamma$ in the 
particle-hole amplitude. Similar to the case of disordered electrons in
zero magnetic field \cite{AA} the first interaction corrections to conductivity
(Altshuler-Aronov type-terms) 
arise from exchange diagrams of the Fig.1 analogous to those of Fig.2
from the second reference in \cite{AA}.

Like in the $B=0$ case \cite{AA} three first diagrams
representing self-energy, vertex and impurity 
scattering rate corrections cancel each other in the leading approximation independently of the nature of interaction. At finite $T<<{1/\tau_{tr}}$ the residual 
terms representing inelastic scattering from thermal gauge
fluctuations give power-law corrections which scale as $T^{3/2}$
in the short range case and as $T^2$ in the Coulomb one in accordance with predictions made in \cite{HLR}.

To estimate the contributions of the last two diagrams of Fig.1 one has to notice an important
 difference between
impurity renormalization of scalar (density) and vector (current) vertices coupled by the gauge
propagator $D_{\mu\nu}(\omega,q)$. 

Each scalar vertex $\Lambda^{(0)}=1$ gets dressed by impurity ladder which 
introduces an extra diffusion propagator into the integrand
$\Lambda(\omega, q)={1\over \tau}
{1\over {Dq^2 -i\omega}}$ \cite{F}.
Then exchange by the scalar component of the gauge propagator $D_{00}(\omega,q)={2\pi\over m}{Dq^2 - i\omega\over Dq^2}$
gives the same contribution as in the case of zero field
$\delta\sigma^{CF}_{sc, exch}={e^2\over \pi h}\log T\tau_{tr}$ \cite{AA,LR}.
A more complete analysis which accounts Hartree terms in addition to the 
exchange contribution yields \cite{CCLM}:
\begin{equation}
\delta\sigma^{CF}_{sc}={e^2\over \pi h}(2-2\log 2)\log T\tau_{tr}
\end{equation} 
On the contrary, a transverse 
vector vertex ${\vec \Lambda}^{(0)}={\vec p \over m}$
acquires only an additional factor ${\tau_{tr}\over \tau}$ \cite{F} while a diffusion pole appears in the longitudinal part proportional
to $\vec q$ which does not contribute to gauge invariant observables. 
Then one can easily see that the contribution of the off-diagonal (Chern-Simons) component $D_{01}(\omega, q)$ vanishes. It should be noted, however, that
at finite residual field $\Delta B$ an exchange by the Chern-Simons component provides a mechanism of additional skew scattering which modifies both longitudinal and Hall conductivities of CF
in the vicinity of $\nu=1/2$ \cite{DVK}.
 
To calculate the contribution coming from transverse gauge fluctuations 
governed by $D_{11}(\omega,q)={1\over {-i\gamma_q \omega + \chi_q q^2}}$ we first do the momentum sum in the CF Green functions.
This sum factorizes into two parts on either side of the impurity ladder.
Each of the factors is proportional to 
$M_{ij}=\sum_{\vec p}G^2_{+}(\vec p)G^2_{-}(\vec p){p_i p_j\over m^2}{\tau_{tr}\over \tau}=i\epsilon^{2}_{F}\tau_{tr}\tau$.
 
In the case of $T=0$ and finite external frequency
$\Omega$ two last diagrams of Fig.1 (plus two similar diagrams with the gauge dressing of the particle and the hole line interchanged) combine into
\begin{equation}
\delta\sigma^{CF}_{vec}(\Omega)=
{ie^2\over 2\pi}
\int^{1/\tau_{tr}}_{\Omega}
{d\omega\over 2\pi}
\int{d{\vec q}\over (2\pi)^2} 
{M_{ij}M^{*}_{ki}(\delta_{ij}-{q_i q_j\over q^2})
\over
 {m\tau^2 (Dq^2 -i(\omega+\Omega))
(-i\omega \gamma_q + {\chi^{\prime}_q} q^2 ) } }
\end{equation}
where ${\chi^{\prime}_q} = \chi_q + {1\over 8\pi m}$. Notice that impurity scattering lifetime $\tau$ drops out of the formula (5).
At finite temperature $T>>\Omega$ the calculation of (5) can be done
by using imaginary frequencies and analytic continuation.
At $k_F l>>1$ and $T<<1/\tau_{tr}$ the integral in (5) can be estimated 
with logarithmic accuracy.  In the case of short-range interaction
we obtain 
\begin{equation}
\delta\sigma^{CF}_{vec}(T)={e^2\over 2\pi h}(\log {T\tau_{tr}})( \log {k_F l})
\end{equation}
As compared to the conventional scalar contribution (4) the $\log T$
correction (6) appears to be enhanced by a large factor $\log k_{F}l$.
In contrast to the conventional Coulomb problem at zero field \cite{AA} the static limit
of the transverse gauge interaction remains singular $(D_{11}(0,q)\sim 1/q^2)$
and  
Hartree terms do not reduce the result in the leading approximation.
 
The negative transverse gauge field contribution (6) diverges as T tends to zero and
manifests a breakdown of perturbation theory at $T\sim 1/\tau_{tr}\exp(-{\pi k_F l\over \log k_F l})$ suggesting 
a localization length $L_{loc}\sim l \exp{{\pi k_F l\over \log k_F l}}$
which is shorter than implied by the scalar term (4) only.

In the unscreened Coulomb case we obtain the result containing double-logarithmic terms 
\begin{equation}
\delta\sigma^{CF}_{vec}(T)={e^2\over 2\pi h}(\log {T\tau_{tr}}) [
\log {k_F l}+{1\over 4}\log {T\tau_{tr}}]
\end{equation}
which holds at $T>T_0 \sim\epsilon_{F}/(k_{F}l)^3$. In the range of temperatures
$T_0 <T<1/\tau_{tr}$ the coefficient in square brackets reduces by a factor of two.
Because of a singular behavior of $D_{11}(0,q)\sim 1/q$ the leading logarithms
in (7) are not affected by Hartree terms either.

At $T<T_{0}$ the divergency in (7) is cut off but at 
temperatures of order $T_{cr}={\kappa^2\over m} k_F l$ one should already expect a crossover to a short-range regime
due to screening by a ground plate placed on distance $\sim \kappa^{-1}$
apart from 2D electron gas.
 
Remarkably, in both cases logarithmic corrections are non-universal
although only weakly dependent on
details of interaction potential $V_q$. In fact, they only depend on $k_F$
and $l$ which can be both extracted from experimental data on surface acoustic wave propagation \cite{exp1}. 

In agreement with general expectations the correction $\delta\sigma^{CF}_{vec}(T)$ is stronger in the short-range case when effects of gauge forces
which physically correspond to local electron density fluctuations
are more pronounced \cite{HLR}.

In terms of physical observables negative corrections
(4) and (6-7) to $\sigma^{CF}(T)$ increase both diagonal conductivity $\sigma_{xx}(T)\approx ({e^2\over 2h})^2 (\sigma^{CF}_{0}+\delta\sigma^{CF}(T))^{-1}$
and resistivity $\rho_{xx}(T)=(\sigma^{CF}_{0}+\delta\sigma^{CF}(T))^{-1}$ 
at half filled Landau level at temperatures $T<1/\tau_{tr}$.
Using typical values of $\epsilon_{F}\sim 20K$ and $k_F l\sim 15$ one can estimate the bare transport scattering rate  $1/\tau_{tr}$ to be of order of $1K$.
It agrees with the threshold temperature below which strong although sample-dependent $\log T$ corrections were 
experimentally observed in \cite{log}. 

In presence of a thermal gradient gauge field corrections to other kinetic coefficients
can be found in a similar way. In particular we expect an enhanced nonlinear
correction to the Seebeck coefficient $\delta S_{xx}(T)\sim T (\log T\tau_{tr})(\log k_F l)$.
The existing experimental techniques \cite{TEP}
should in principal allow observation of
such an effect at temperatures below $100 mK$ where corrections from
phonon scattering of CF varying as $T^3$ become negligible. 
A detailed discussion of effects of transverse gauge fluctuations on thermoelectric transport properties will be presented elsewhere \cite{DVK}.

We also note that a more complete analysis of gauge interactions beyond the
lowest order can be achieved by means of a renormalization group 
study of the scale dependent conductivity $\sigma_{xx}(L)$
\cite{CCLM}. 
It is not clear at this point if double-logarithmic terms appearing in the 
Coulomb case spoil renormalizability of the noninteracting theory 
of fermions in RMF. 
The work in this direction is in progress.

In conclusion, 
we propose an explanation of recently observed strong non-universal
$\log T$ dependence of the conductivity at half filled Landau level.
A qualitative agreement between the theory and experimental data
\cite{log} provides a further support for the idea of CF and their transverse gauge interactions.

pagebreak

\end{document}